\documentclass[12pt,manuscript]{emulateapj}%{aastex}%

\shorttitle{Visual orbit of GJ 164 AB}
\shortauthors{F. Martinache et al.}

\begin{document}

\bibliographystyle{apj}

\title{VISUAL ORBIT OF THE LOW-MASS BINARY GJ 164 AB}

\author{Frantz Martinache\altaffilmark{1},
  Barbara Rojas-Ayala\altaffilmark{1},
  Michael J. Ireland\altaffilmark{2},
  James P. Lloyd\altaffilmark{1} and
  Peter G. Tuthill\altaffilmark{2}}

\altaffiltext{1}{Department of Astronomy, Cornell University, Ithaca NY}
\altaffiltext{2}{School of Physics, University of Sydney, Sydney NSW
  Australia}

\begin{abstract}
We report seven successful observations of the astrometric binary GJ
164 AB system with aperture masking interferometry. The companion,
with a near infrared contrast of 5:1 was detected beyond the formal
diffraction limit.
Combined with astrometric observations from the literature, these
observations fix the parallax of the system, and allow a
model-independent mass determination of both components.
We find the mass of GJ 164B to be $ 0.086 \pm 0.007 M_{\sun}$.
An infrared spectroscopic study of a sample of M-Dwarfs outlines a
method for calibrating metallicity of M-Dwarfs.
Results from the newly commissionned TripleSpec spectrograph reveal
that the GJ 164 system is at least of Solar metallicity. Models are
not consistent with color and mass, requiring a very young age to
accommodate a secondary too luminous, a scenario ruled out by the
kinematics.
\end{abstract}


\section{INTRODUCTION}

The development of a self-consistent theory of the internal structure
of stars and the construction of models which trace evolutionary
behaviour are major achievements of modern astrophysics. Both theory
and models have however long been centered on intermediate and high
mass stars, with little consideration of objects with masses below 0.6
M$_{\sun}$.
The realization that the Solar neighborhood is overwhelmingly
dominated by low mass stars \citep{1971ARA&A...9..103V,
  1998ASPC..134...28H}, and the discovery of brown dwarfs
\citep{1995Natur.378..463N} have brought a lot more attention to the
lower end of the main sequence and what lies below.

The complex physical nature of these very low mass objects makes
their modeling intricate. Boundaries between stars, brown dwarfs and
planets, as well as theoretical relations that predict the luminosity
and temperature as functions of age, mass and metallicity are still
largely untested in the relevant range of parameters.
Correct calibration of these models is of extreme importance, since
they are now extrapolated to estimate the mass of brown dwarfs
\citep{2006ApJ...640.1063B} and giant exoplanets
\citep{2003A&A...402..701B} from their luminosity and age.

The observations required to challenge and improve the models are
dynamical mass measurements of multiple star systems, combined with
accurate photometry and distance determination. 
Besides gravitational microlensing \citep{1986ApJ...304....1P},
the observation of binary systems and the use of Newtonian orbital
dynamics provide the only method of directly measuring accurate
stellar masses.
With maximum sensitivity at separations greater than 1 arcsec,
conventional Adaptive Optics (AO) observations at Palomar are limited
to the most nearby objects or the consequent orbital periods become
too long to lead to dynamical masses.
We have therefore been using aperture masking interferometry
\citep{2000PASP..112..555T} in conjunction with AO
\citep{2006ApJ...650L.131L}.
The precision calibration of the data achieved with this observing
mode indeed leads to reliable results up to and beyond the formal
diffraction limit (super resolution), and very precise photometry.
Combined with radial velocity \citep{2007ApJ...661..496M} or
astrometry \citep{2008ApJ...678..463I}, aperture masking
interferometry has so far provided some of the most precise
(dynamical) masses of objects below 0.1 M$_{\sun}$.

\section{CHARACTERIZATION OF THE ORBIT}

We report here aperture masking interferometry observations of the
astrometric binary GJ 164 AB.
GJ 164 (aka LHS 1642, G 175-19 or Ross 28) is a high proper motion
star, catalogued as a M4.5 Dwarf \citep{1995AJ....110.1838R},
11-13 pc distant.
From the 2MASS catalog, its apparent near infrared magnitudes are
$J=8.773 \pm 0.032$, $H=8.248 \pm 0.030 $, and $K=7.915 \pm 0.016$.
The discovery of GJ 164 B was reported by \citet{2004ApJ...617.1323P},
as a part of the STEPS program of \citet{1996ApJ...465..264P}.
The combination of the astrometry data from the discovery paper,
combined with the aperture masking interferometry data reported in
this paper provides a complete dynamical characterization of the
system, resulting in masses that are independent of the use of a
model or a mass-luminosity relation.
GJ 164 is an especially interesting target for the mass of the B
component lies very close to the substellar limit.

\subsection{Aperture Masking Interferometry}
\label{sec:apm}

Measuring the dynamical masses of a low-mass binary system requires
some patience, for the observations have to cover at the very least
a significant fraction of the orbital period.
Because of its low mass ($<$ 0.6 M$_{\sun}$), a relatively short period
($<$ 10 yr) binary will have a semimajor axis smaller than 4 AU. Even
for objects as close as 10 parsecs, the angular separation won't
exceed 0.4 arcsecs, a performance theoretically within the grasp of a
5-10 meter class telescope equipped with AO.
Yet in practice, the residual quasi-static speckles in the AO PSF halo
seriously limit the capability of detecting a companion at angular
separation smaller than 2-4 $\lambda/D$ \citep{1999PASP..111..587R}.
Aperture masking interferometry with AO adresses this issue.

A non redundant mask discards most of the pupil by sampling a
few spatial frequencies only \citep{1988AJ.....95.1278R,
1989AJ.....97.1510N}. One admittedly loses most of the light
(the 9-hole mask used for this works transmits approximatively 15 \%)
but rejects all atmospheric noise, as well as internal aberrations and
non common path errors, which in turn, dramatically increases the
signal-to-noise ratio \citep{2000PASP..112..555T}.
Used in conjunction, aperture masking interferometry and AO provide
stable fringes, and enable long integration times, therefore making
faint target accessible. The only limit is whether the AO system can
lock on the system. Thus, any target that can be observed with AO can
also be observed with aperture masking interferometry. With a visible
magnitude V = 13.5 \citep{1996AJ....112.2300W}, GJ 164 can fairly
easily be observed with aperture masking.
If calibration is sufficiently accurate, the interferometry enables
super-resolution, ie. the detection of structures on a target at a
scale smaller than $\lambda/D$. The data presented in Tables
\ref{tbl:ao} and \ref{tbl:phot} shows that the companion to GJ 164 A
is detected on multiple occasions in that super-resolution regime.
\citet{2006SPIE.6272E.103T} provide a recent description of
the general principles and performances of aperture masking
interferometry while \citet{2006ApJ...650L.131L} detail the
experiment undertaken at Palomar.

The AO observations performed with PHARO \citep{2001PASP..113..105H}
at Palomar as well as with NIRC2 at Keck II span the range December
2003 to December 2006. The companion was successfully detected all
seven runs.
The data are calibrated with observations of one or several stars of
similar brightness for both the AO wave front sensor and the science
camera to ensure comparable wave front correction and signal-to-noise
ratio.
The final dataproduct of our custom software pipeline written in IDL,
is a collection of calibrated closure phases
\citep{1986Natur.320..595B}.
This very robust observable rejects both atmospheric noise and
calibration errors of the wave front sensor.
The 9 hole mask provides $(^9_3)$=84 possible closure triangles,
$(^8_2)$ = 28 of which are theoretically independent.

\begin{figure}
\center{\resizebox*{\columnwidth}{!}{\includegraphics{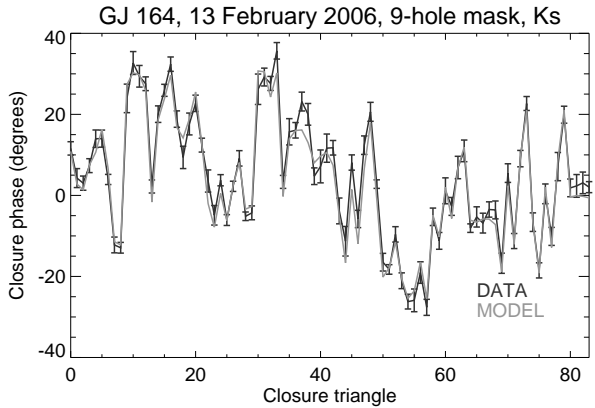}}}
\caption{Calibrated closure phases observed on JD 2453779.71 with a
  K$_S$ filter. A binary star model (light gray) with a 86 mas separation,
  a position angle of 102$^{\circ}$ and a contrast of 5:1 satisactorily
  matches the observations (dark gray).}
\label{fig:cp}
\end{figure}

The instantaneous geometry of a binary star is modeled by a set of
three parameters: contrast ratio $c$, angular separation $a$ and
position angle (PA) $\theta$. The example of one such model fitted to
a set of closure phases is given in Figure \ref{fig:cp}. The aperture
masking data processing software typically uses several hundreds of
frames which are averaged to produce the closure phases. Due to the
central limit theorem, the distribution of closure phases $\{\phi_i\}$
can be considered Gaussian, with uncertainties $\sigma_i$ given by the
standard deviation to the mean. The best fit minimizes the
traditionally used goodness-of-fit parameter $\chi^2$ defined as

\begin{equation}
  \chi^2 \equiv \sum_i\frac{(\phi_i - f_i(a,\theta,c))^2}{\sigma_i^2},
  \label{eq:chi2}
\end{equation}

\noindent
where $f_i$ designates the model. A systematic error term is
added to the closure phase dispersion $\sigma_i$ to achieve
a reduced $\chi^2_{\nu}=1$. The likelihood of the parameters
given the set of closure phases $\{\phi_j\}$ is related to the
$\chi^2$ by

\begin{equation}
  L(a,\theta,c|\{\phi_j\}) \propto \exp (-\chi^2/2).
  \label{eq:likely}
\end{equation}

Normalized, $L$ is the full joint probability density function for all
three parameters. The confidence interval of one individual parameter
is calculated by integrating out the two others. 
For example, the marginal probability density function for the angular
separation is calculated by integrating out the position angle and the
contrast ratio:

\begin{equation}
  p(a) = \int d\theta\,dc\,L(a,\theta,c).
  \label{eq:margin}
\end{equation}

The results of this analysis: relative position and photometry are
gathered in Tables \ref{tbl:ao} and \ref{tbl:phot}.
The more or less favorable seeing conditions explain the variable
confidence interval sizes. We achieve a precision of a few
milliarcseconds at separations as low as 44 mas in K$_S$ band (0.5
$\lambda$/d) on good nights at Palomar.

%% ----------------------------------------
%%                AO table
%% ----------------------------------------
\begin{deluxetable}{lllr@{ $\pm$ }l r @{ $\pm$ } l}
\tablecolumns{8}
\tabletypesize{\scriptsize}
\tablewidth{0pc}
\tablecaption{
  APERTURE MASKING MEASUREMENTS AT PALOMAR AND KECK:
  angular separation and position angle of GJ 164 B.
}
\tablehead{
  \colhead{Julian Date} & \colhead{Band} & \colhead{Telescope} &
  \multicolumn{2}{c}{Sep.} & \multicolumn{2}{c}{PA} \\
  \colhead{(-2,450,000)}
}
\startdata
3004.85 & H        & Palomar & 82.8 & 1.6 & 111.1 & 0.9 \\ % 031229
3632.96 & H        & Palomar & 53.3 & 1.3 & 149.1 & 1.2 \\ % 050918
3723.66 & H        & Palomar & 80.1 & 3.3 & 116.1 & 2.1 \\ % 051218
3779.71 & H, K$_S$ & Palomar & 86.1 & 2.0 & 102.1 & 0.9 \\ % 060213
3957.62 & H        & Keck    & 49.3 & 0.5 &  47.0 & 0.6 \\ % 060810
4018.98 & K$_S$    & Palomar & 39.4 & 1.3 & 344.8 & 2.6 \\ % 061010
4078.79 & K$_S$    & Palomar & 57.3 & 2.2 & 294.1 & 2.8    % 061209
\enddata
\label{tbl:ao}
\end{deluxetable}

%% ----------------------------------------
%%          Photometry table
%% ----------------------------------------
\begin{deluxetable}{ccr@{ $\pm$ }l}
\tablecolumns{4}
\tablewidth{0pc}
\tablecaption{GJ 164 PHOTOMETRY}
\tablehead{
  \colhead{Julian Date} & \colhead{Filter} &
  \multicolumn{2}{c}{Contrast} \\
  \colhead{(-2,450,000)}
}
\startdata
3004.85 & H     & 5.97 & 0.32 \\
3632.96 & H     & 5.77 & 0.46 \\
3723.67 & H     & 5.78 & 0.63 \\
3779.71 & H     & 5.53 & 0.60 \\
        & K$_S$ & 4.89 & 0.19 \\
3957.62 & H     & 5.42 & 0.03 \\
4018.98 & K$_S$ & 8.66 & 4.81 \\
4078.79 & K$_S$ & 5.67 & 1.17
\enddata
\label{tbl:phot}
\end{deluxetable}

\subsection{Astrometry}
\label{sec:astrometry}

GJ 164 was discovered as a binary in an astrometric survey by
\citet{1996ApJ...465..264P, 2004ApJ...617.1323P}.
While the interferometry resolves the binary and provides
its instant geometric configuration, the astrometry records
the position of a star's photocenter, measured relative to
several more distant stars. Repeated observations reveal the
star's proper motion as well as its parallax.
Once these two effects substracted from the astrometric
signal, any residual wobble of the photocenter can be
attributed to the presence of one or more companions,
massive enough to cause an observable shift in position.
The orbit of the companions can then be fully characterized.

Astrometric data were extracted from the figures of
\citet{2004ApJ...617.1323P} then added the motion due to the proper
motion and the parallax given in the paper's tables. These extracted
values are gathered in Table \ref{tbl:steps}. We initially assume a
2.0 mas uncertainty for each individual measurement, as suggested by
the residuals of the analysis performed by
\citet{2004ApJ...617.1323P}.

%% ----------------------------------------
%%          Astrometry table
%% ----------------------------------------
\begin{deluxetable}{crr}
\tablecolumns{3}
\tablewidth{0pc}
\tablecaption{GJ 164 ASTROMETRY}
\tablehead{
  \colhead{Julian Date} & \colhead{$\Delta$R.A.} &
  \colhead{$\Delta$ Decl.} \\
  \colhead{(-2,450,000)}
}
\startdata
0801.5      &     0.00  &     0.00 \\
0801.5      &     0.00  &    -2.80 \\
1088.3      &  -147.91  &  -646.96 \\
1189.0      &  -328.82  &  -874.15 \\
1189.0      &  -324.82  &  -874.05 \\
1189.0      &  -325.32  &  -871.65 \\
1436.5      &  -457.34  & -1409.82 \\
1494.0      &  -563.37  & -1519.60 \\
2657.5      & -1639.04  & -4132.08 \\
2889.5      & -1745.18  & -4629.59 \\
2889.5      & -1745.18  & -4627.29 \\
2889.5      & -1745.88  & -4625.69 \\
3030.5      & -2014.31  & -4953.05 \\
3030.5      & -2014.61  & -4957.45 \\
3051.5      & -2044.87  & -5016.85
\enddata
\label{tbl:steps}
\end{deluxetable}

The astrometric residual wobble however remains a function of the
fractional light of the secondary, $\beta=L_2/(L_1+L_2)$, with $L_1$
and $L_2$ respectively denoting the luminosity of the primary and the
secondary, as well as the fractional mass of the secondary
$f=m_2/(m_1+m_2)$.
In the limit at which the light of the secondary is negligible,
the photocentric orbit is identical to the Keplerian orbit of
the primary around the actual center of mass of the system.
If the luminosity of the secondary is not so negligible, then the
photocentric semimajor axis $\alpha$ is reduced proportionally to
$\beta$. The ratio of the photocentric orbit $\alpha$, and the
Keplerian orbit $a$ is \citep{1988ApJ...333..943M}:

\begin{equation}
\alpha/a = f - \beta.
\label{eq:astrometry}
\end{equation}

A relatively small-mass, non-luminous secondary, and a relatively
large-mass, luminous secondary will therefore have indentical
astrometric signatures. The data presented in Section \ref{sec:apm}
however rules out the latter possibility. Indeed, the aperture masking
interferometry data gathered in Table \ref{tbl:phot} provides the H
and K$_S$ band contrast ratios, which are turned into the following
differences of magnitude:

\begin{eqnarray*}
  \Delta H   &=& 1.835 \pm 0.006 \\ % Keck datapoint
  \Delta K_S &=& 1.721 \pm 0.097.   % Best Palomar datapoint
\end{eqnarray*}

These contrast ratios can be used to decompose the 2MASS combined
magnitudes of the binary into apparent magnitudes for individual
components:
$H = 8.432 \pm 0.030$  and $K = 8.117 \pm 0.017$ for GJ 164A,
$H = 10.267 \pm 0.030$ and $K = 9.834 \pm 0.038$ for GJ 164B.

The corresponding (H-K) color indices:
$(H-K)_A = 0.315 \pm 0.034$ and $(H-K)_B = 0.433 \pm 0.048$ alone,
provide a parallax-independent observable. While the $(H-K)$ color
index of the primary is well compatible with a M4.5 spectral type, the
secondary is very unlikely to be any earlier than M8 (see for instance
\citet{1983A&A...128...84K}). A well characterized nearby equivalent
is the star VB 10 from the 8 parsec survey \citep{1994AJ....108.1437H}.

We use VB 10 as a standard to estimate the contrast ratio of the two
components of GJ 164 in the R band where the astrometry data was
taken. The 2MASS catalogue provides a $(R-K)$ = 6.8 color index for
VB 10, which combined with the data gives an apparent $R = 16.6$
magnitude for GJ 164B. \citet{2003AJ....125..984M} give $R = 12.4$
for GJ 164A, which corresponds to a contrast ratio $c \approx 47$.
This gives an upper limit to the fractional light of the secondary:
$\beta = 0.021$, which means that the hypothesis of a secondary of
negligible visible luminosity would, at most, underestimate the mass
ratio (cf. eq. \ref{eq:astrometry}) by 2.1 \%.
This systematic uncertainty needs to be added up to the statistical
uncertainty deduced from the $\chi^2$ analysis performed in Section
\ref{sec:dynaMass}.

\section{CHARACTERISTICS OF GJ 164 AB}

\subsection{Orbital Parameters}

\label{sec:anal}

\begin{figure}
\center{\resizebox*{
    \columnwidth}{!}{\includegraphics{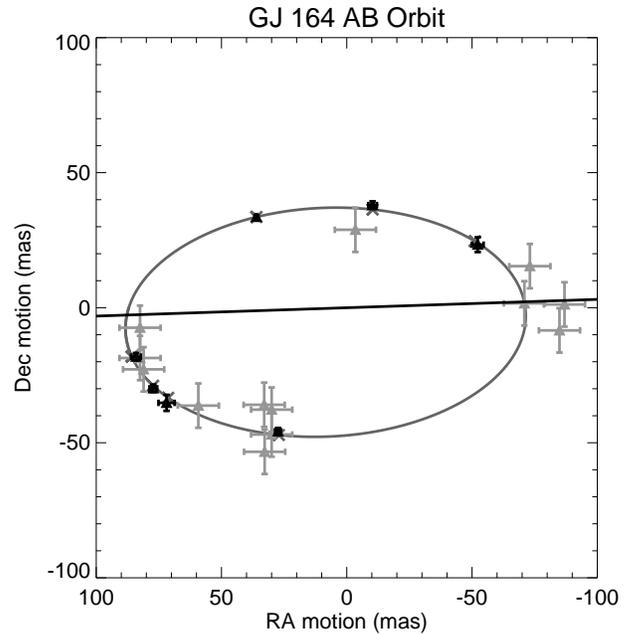}}}
\caption{GJ 164 B ORBIT. The measurements and associated uncertainties
  of the Aperture Masking data are represented by dark gray
  rectangles. 
  Light gray triangles represent the STEPS astrometry data. The
  relative apparent orbit (curve) is clockwise. Table \ref{tbl:orb_els}
  provides the corresponding orbital elements. The almost horizontal
  straight line represents the line of nodes.}
\label{fig:orbit}
\end{figure}

The aperture masking observations gathered in Table \ref{tbl:ao}
provide the instant position of the companion to the primary. These
observations alone suffice to constrain all seven orbital elements
of the binary, independently of the parallax, with the semimajor axis
expressed in angular units.

We use a model identical to the one described in Section 3.1.2 of
\citet{2007ApJ...661..496M} to fit a seven-parameter model to 14
observables. The observables are two coordinates for seven aperture
masking data points. The seven orbital parameters are:
the angular semimajor axis $\alpha$, the eccentricity $e$, the
longitude of the ascending node $\omega_0$, the inclination $i$,
the argument of the periastron $\Omega_1$, the period $P$ and the
epoch at periastron passage $T_P$.
Our solution exhibits a final reduced $\chi^2_{\nu} = 0.61$ for 7
degrees of freedom.
Note that the temporal coverage of these observations definitely rules
out the possibility of a $\sim$ 4 yr orbital period suggested by
\citet{2004ApJ...617.1323P}. The seven orbital parameters and the
associated uncertainties are presented in Table \ref{tbl:orb_els}.

We then proceed to a joined fit AO + astrometry, that is a
thirteen-parameter fit to 40 observables. The extra parameters
included in this new analysis are: the two components of the
proper motion, the relative parallax $\Pi$ as well as the semimajor
axis of the motion of the primary around the center of mass of the
system $a_1$. With a 2 mas uncertainty on the individual astrometric
measurements, as assumed in \citet{2004ApJ...617.1323P}, the solution
to this global fit exhibits a final reduced $\chi^2_{\nu} = 0.73$ for
27 degrees of freedom. We therefore assumed astrometric errors of 1.4
mas so that the reduced $\chi^2_{\nu}$ is exactly 1.
Table \ref{tbl:orb_els} compares the seven orbital parameters and
their uncertainties obtained from the fit of the AO data only and the
joined fit: both approaches produce compatible solutions. Figure
\ref{fig:orbit} represents the orbit resulting from this joined fit,
onto which are superposed the AO and astrometry data.

\subsection{Dynamical Masses}
\label{sec:dynaMass}

Like \citet{2004ApJ...617.1323P}, we use a $2 \pm 1$ mas correction to
convert the relative parallax $\Pi = 70.0 \pm 0.8$ mas, to the absolute
parallax: $\Pi = 72.0 \pm 1.2$ mas.
We can now determine the actual relative semimajor axis of the binary:
$a = 1.1 \pm 0.2$ AU, as well as the total mass:

\begin{equation}
  M_T = a^3/P^2,
\end{equation}

\noindent
that is $M_T = 0.343 \pm 0.026\,M_{\sun}$. In the limiting case where
the luminosity of the secondary is negligible, the fractional mass of
the secondary (cf. eq. \ref{eq:astrometry}) is given by the ratio of
the photocentric and relative orbit semimajor axis, that is $f = 0.250
\pm 0.010$.
It is remarkable that the fractional mass should be a ratio of two
angular radii, for this makes this quantity parallax-independent.
Section \ref{sec:astrometry} shows there is at least a 3.8 magnitude
difference between the two components of the binary, which translates
into a fractional light $\beta = 0.021$. The fractional mass $f$
therefore requires non symetric uncertainties: 
$f = 0.250_{-0.010}^{+0.031}$.

From the total mass and the fractional mass, one can infer the masses
of both components: $M_2 = 0.086_{-0.007}^{+0.012}$ and
$M_1 = 0.257_{-0.022}^{+0.020}$. % these are checked.
All dynamical characteristics of the GJ 164 system are given in Table
\ref{tbl:dynaMass}.

%I have to say I am a bit disappointed. Although the orbit is pretty
%good, the precision on the mass is not so great, with at best a 10\,\%
%uncertainty on GJ 164 B.

%Why is that not as conclusive as what we did with GJ 623? Well, the
%uncertainties on the parameters are of the same order of magnitude,
%which means that the data is of comparable quality (the AO data, that
%is...). But the orbital period is about half of GJ 623, same thing for
%the parallax, and the semimajor axis is about four times smaller than
%the one of GJ 623. The resulting relative errors, which are what
%matter in the end for the determination of the uncertainties on the
%mass, are in return twice or four times bigger.

%Should we be happy with what we have here? I'd say yes... but what do
%we do from here?

%% ----------------------------------------------------------------
%%       ORBITAL ELEMENTS (a, e, i, W1, w0, P, Tp)
%% ----------------------------------------------------------------
\begin{deluxetable}{l  c @{ $\pm$ } l  c@{ $\pm$ } l }
\tablecolumns{5}
\tablewidth{0pc}
\tablecaption{ORBITAL ELEMENTS}
\tablehead{
  \colhead{Parameter} & \multicolumn{2}{c}{AO only} &
  \multicolumn{2}{c}{AO + STEPS}
}
\startdata
$\alpha$ (mas)     &  80.4   & 1.3   & 80.5   & 1.2   \\
$e$                &  0.157  & 0.012 & 0.161  & 0.012 \\
$i$ (deg)          &  121.9  & 0.9   & 121.9  & 0.8   \\
$\Omega_1$ (deg)   &  272.4  & 1.3   & 271.8  & 1.2   \\
$\omega_o$ (deg)   &  314.7  & 5.0   & 311.7  & 3.7   \\
$P$ (days)         &  734.3  & 4.6   & 736.9  & 1.7   \\
$T_P$ (reduced JD) &  1868   & 19    & 1856   & 8   
\enddata
\label{tbl:orb_els}
\end{deluxetable}

%% ----------------------------------------------------------------
%% ASTROMETRIC ELEMENTS (pm, )
%% ----------------------------------------------------------------
\begin{deluxetable}{l  r @{ $\pm$ } l}
\tablecolumns{3}
\tablewidth{0pc}
\tablecaption{ASTROMETRIC ELEMENTS}
\tablehead{
  \colhead{Parameter} & \multicolumn{2}{c}{Value}
}
\startdata
Proper motion R.A.  (mas) & -324.5 & 0.3 \\
Proper motion Decl. (mas) & -808.3 & 0.3 \\
Relative Parallax (mas)   &  70.0  & 0.8 \\
Phot. semimaj. axis (mas) &  20.1  & 0.8 \\
Total Mass   (M$_{\sun}$)  & 0.343   & 0.026
\enddata
\label{tbl:ast_els}
\end{deluxetable}

%% ----------------------------------------------------------------
%%                     DYNAMICAL MASSES
%% ----------------------------------------------------------------
\begin{deluxetable}{l c  @{ = } r @{ $\pm$ } l}
\tablecolumns{4}
\tablewidth{0pc}
\tablecaption{DYNAMICAL MASSES: obtained from the joined analysis
  AO+STEPS. See text for more details on uncertainties!}
\tablehead{
  \colhead{Parameter} & \multicolumn{3}{c}{Value}
}
\startdata
Total Mass     & $M_{T} $ & 0.343 &  0.026 $M_{\sun}$ \\
Primary Mass   & $M_{1} $ & 0.257 &  0.020 $M_{\sun}$ \\
Secondary Mass & $M_{2} $ & 0.086 &  0.007 $M_{\sun}$ \\
Mass ratio     & $M_2/M_T$ & 0.250 & 0.010           
\enddata
\label{tbl:dynaMass}
\end{deluxetable}

\section{SPECTROSCOPY}
\label{sec:spectro}

Careful characterization of the atmosphere and determination of
abundances is necessary to understand the location of GJ 164 AB in a
mass-luminosity (M/L) diagram, relative to an observational M/L
relation or to numerical simulations of M-Dwarf atmospheres.
Determining the overall metallicity of mid-to-late lype M-Dwarfs is
however a delicate task. Indeed, with (V-K) $>$ 5 for spectral types
later than M-4, optical spectroscopy doesn't appear as an efficient
approach, while in the infrared, there is no well-established proxy
for metallicity that calibrates with its optical counterpart.

One strategy to solve this problem is to observe binary systems where
the primary is a Solar-type star (F,G or K-Dwarf) of known metallicity
and the secondary a M-Dwarf. Assuming that both components formed from
the same original molecular cloud, they should share the same
metallicity.
In the context of addressing the planet-metallicity correlation for
M-Dwarf planet surveys \citep{2007SPIE.6693E..26E, 2008psa..conf..303M}
we have begun a separate study of a population of such objects
\citep{2008coolstars15}. Initial results from this study and its
conclusions for the metallicity of GJ 164 are presented in this
section.

\subsection{Observations and Data Reduction}

Our spectroscopic sample consists of five M4-5-Dwarfs (in complement
to GJ 164 AB) associated to Solar type stars, whose metallicity was 
measured by \citet{2005ApJS..159..141V}. Their designation, spectral
type, V and K apparent magnitudes, as well as metallicities are listed
in Table \ref{tbl:sample}.

%% ----------------------------------------------------------------
%%                     THE SAMPLE
%% ----------------------------------------------------------------

\begin{deluxetable}{l l l l r @{ $\pm$ } l r} % <--- column justification (center/left/right)
\tablecolumns{6}
\tabletypesize{\scriptsize}
\tablecaption{The Sample}
\tablewidth{0pc}
\tablehead{   % column headings
  \colhead{Name} &
  \colhead{Sp. Typ.} &
  \colhead{V} &
  \colhead{K} &
   \multicolumn{2}{c}{[M$/$H]} &\\
 & &\colhead{[mag]} &\colhead{[mag]}&\multicolumn{2}{c}{dex}
}
\startdata
GJ 324 B        & M4    & 13.15 & 7.67 & +0.25 & 0.03\\  %signals new line J=8.56
GJ 544 B        & M4    & 14.50 & 9.59 & -0.15 & 0.03\\ % 10.49
GJ 611 B        & M4    & 14.20 & 9.16 & -0.49 & 0.03\\ %9.90
GJ 166 C        & M4.5  & 11.17 & 5.96 & -0.08 & 0.03 \\%6.75
NLTT 25869      & M5    & 14.50 & 8.64 & +0.27 & 0.03 %9.51
\enddata
%\tablenotetext{}{Notes.- V-band photometry is from SIMBAD database
%and K-band photometry is from the 2MASS Point Source Catalog.} 

 \label{tbl:sample}
\end{deluxetable}

\begin{figure*}%[!h]
\center{\resizebox*{\textwidth}{!}{\includegraphics{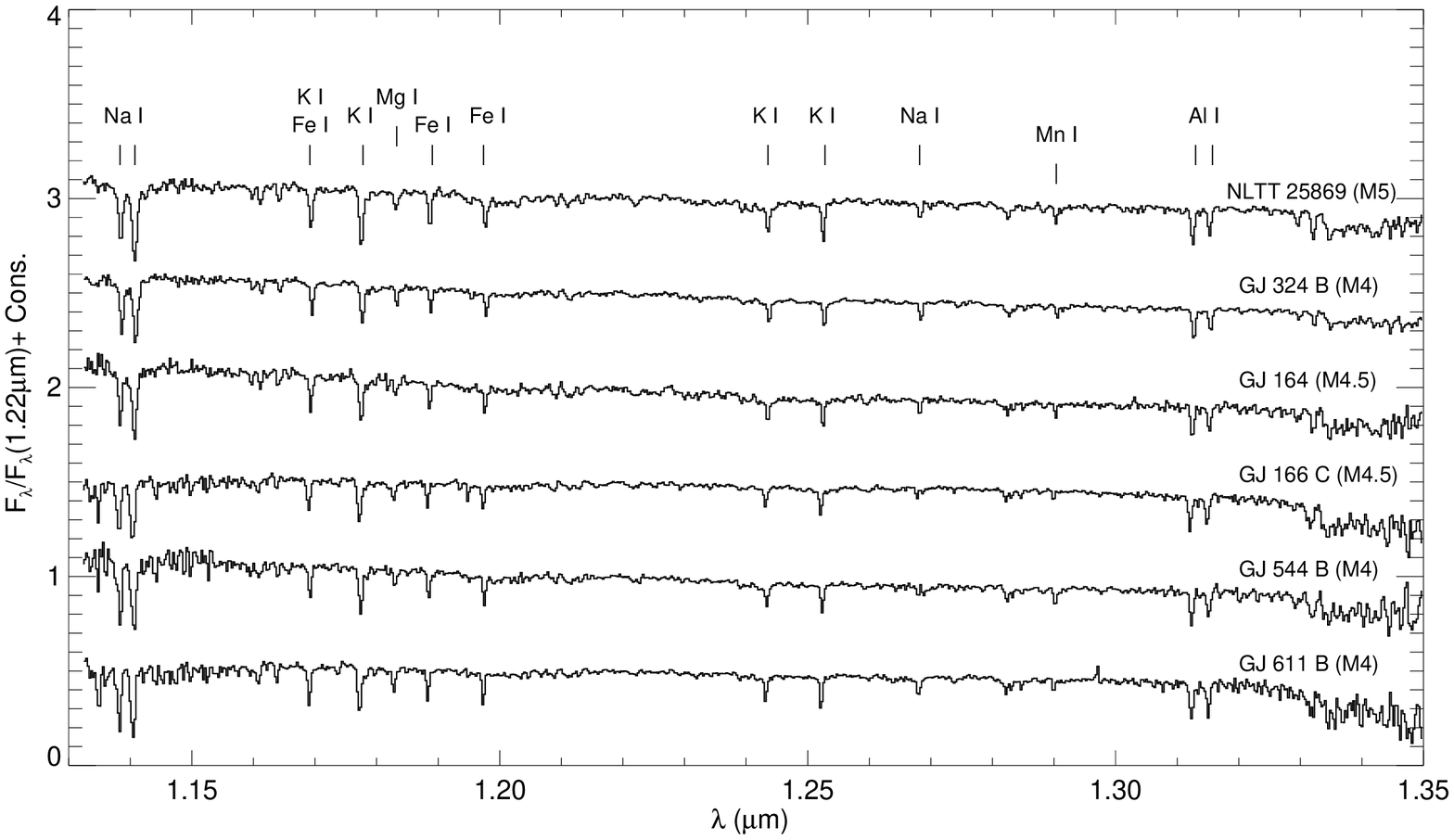}}}
\caption{J-band spectra of the stars ordered from top to bottom by
  decreasing metallicity (listed in Table   \ref{tbl:sample}).
  The most prominent spectral features are highlighted.}
\label{fig:spectrumJ}
\end{figure*}

\begin{figure*}%[!h]
\center{\resizebox*{\textwidth}{!}{\includegraphics{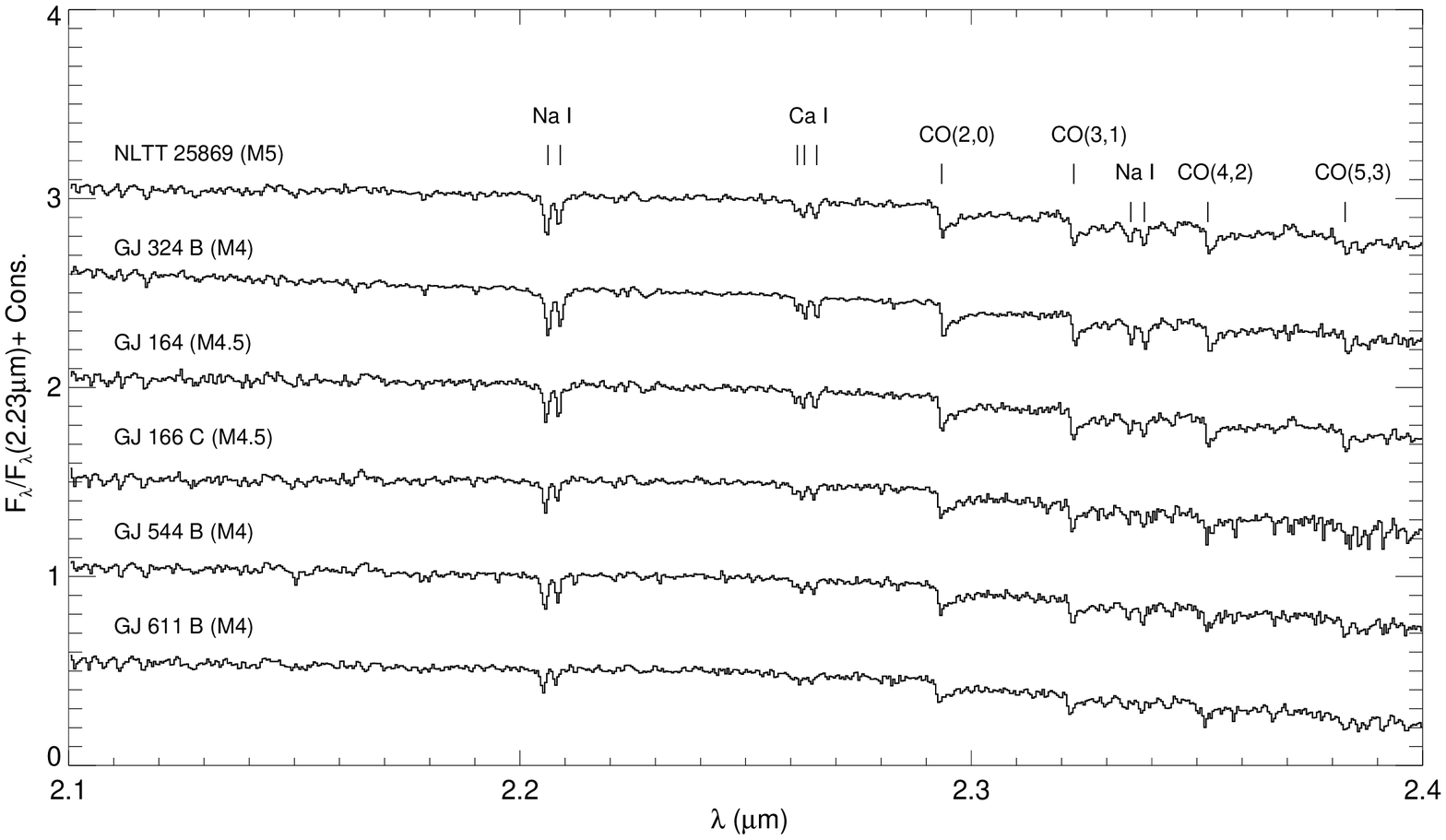}}}
\caption{K-band spectra of the stars ordered from top to bottom by
  decreasing metallicity (listed in Table \ref{tbl:sample}).
  The most prominent spectral features are highlited.}
\label{fig:spectrumK}
\end{figure*}

\begin{figure}
\center{\resizebox*{
    \columnwidth}{!}{\includegraphics[scale=0.7]{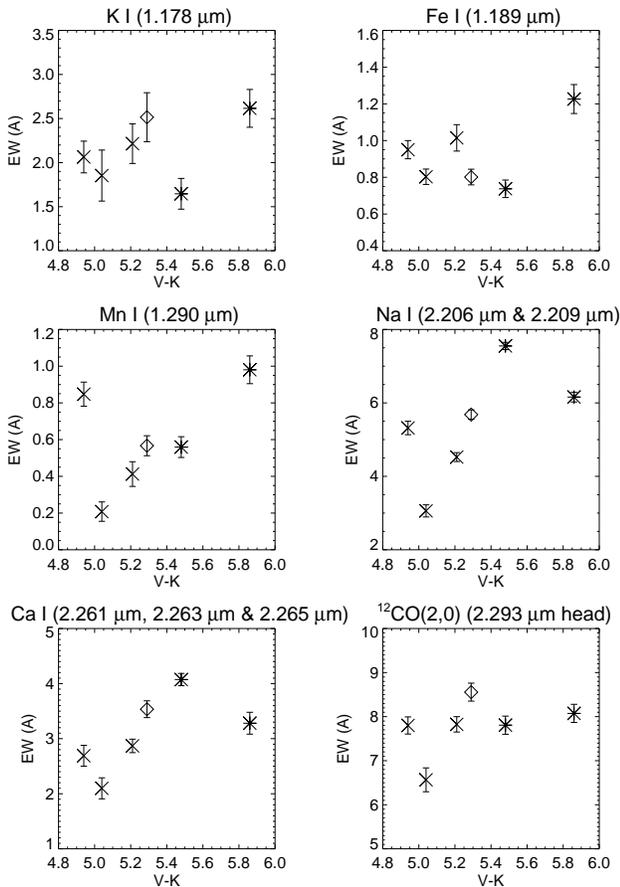}}}
%\center{{!}{\includegraphics[scale=0.7]{./figs/ews_4.ps}}}
%\center{\resizebox*{\textwidth}{!}{\includegraphics{./figs/ews_4.ps}}}
\caption{Equivalent widths of some of the most prominent lines in the
  sample spectra with 1 $\sigma$ errors versus (V-K) colors. A diamond
  represents GJ 164 AB, metal-rich objects  ([M$/$H] $>$ 0.0) are
  represented by asterisks and metal-poor objects ([M$/$H] $<$ 0.0) by
  crosses.}
\label{fig:ews}
\end{figure}

\begin{figure}
\center{\resizebox*{
    \columnwidth}{!}{\includegraphics{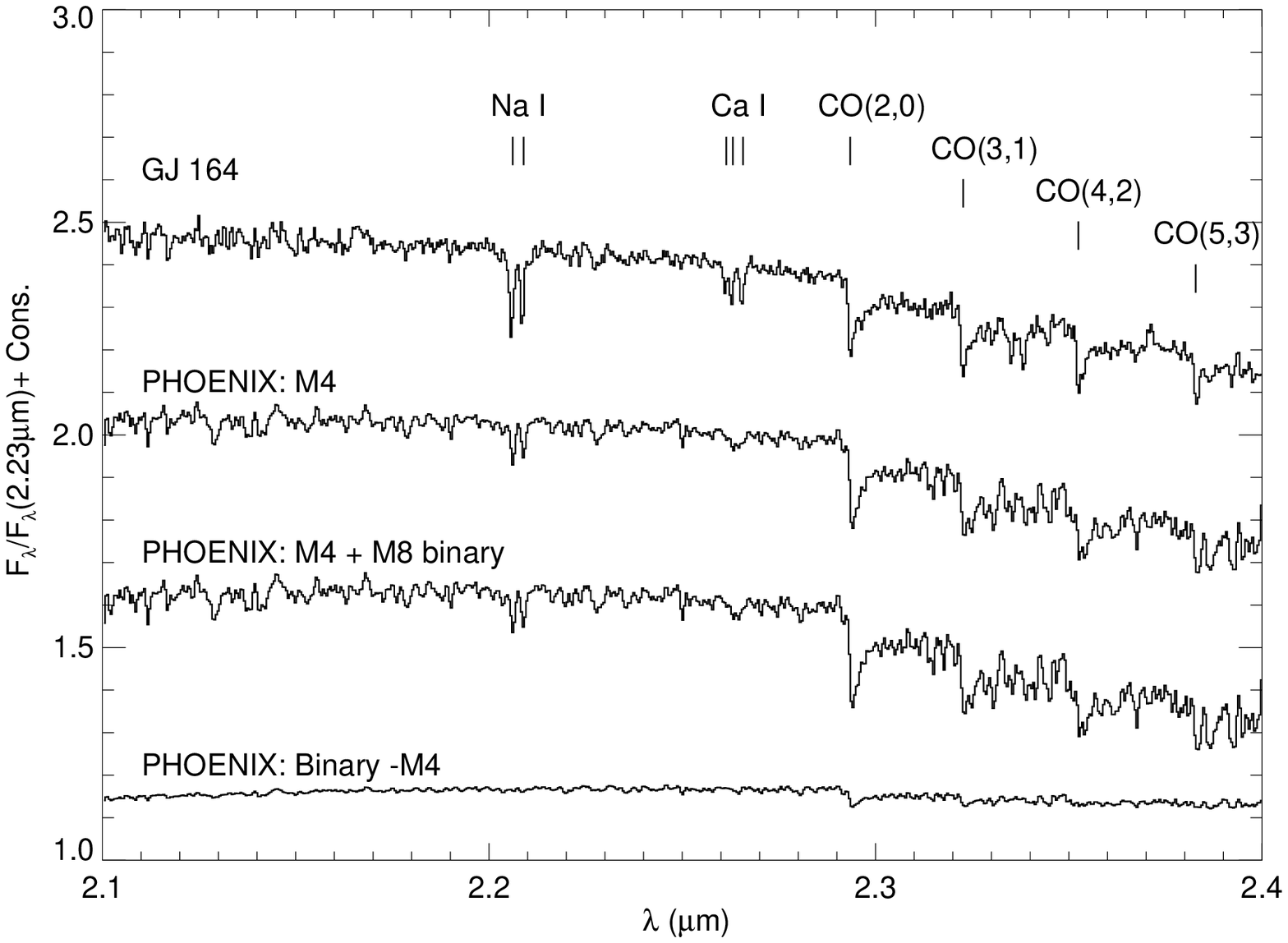}}}
\caption{
  Spectrum of GJ 164 (top) compared to the {\scriptsize PHOENIX}
  models of \citet{1999ApJ...512..377H} for Solar metallicity:
  the models (labeled {\scriptsize PHOENIX:M4}) reproduce most
  features of the spectrum. The presence of a M8-Dwarf companion with
  a 5:1 K-band contrast ratio (labeled {\scriptsize PHOENIX:M4+M8})
  only modifies the strength of the CO bands, as shown in the plot of
  the difference between the two models (bottom).}
\label{fig:mod_diff}
\end{figure}

Near-infrared spectra of these objects were obtained with the recently
commissioned TripleSpec spectrograph on the Palomar Hale Telescope
\citep{2004SPIE.5492.1295W,herter:70140X}. TripleSpec at Palomar is a
$\lambda/\Delta\lambda$ $\approx$ 2500-2700 cross-dispersed
near-infrared spectrograph with a broad wavelength coverage across 5 
simultaneous orders (1.0-2.4 $\mu$m) in echelle format. Its entrance
slit is 1x 30 arcseconds and a notable feature is that it has no
moving parts. 

The data were reduced with an IDL-based data reduction pipeline
developed by P. Muirhead for TripleSpec at Palomar. The data were
sky-subtracted using a sky-frame made by median combination of the 5
exposures on different positions along the slit of the object.
To correct for telluric absorption features and flux-calibrate
their spectra, an A0 V star was observed as close to the science
object star airmass as possible.
Each sky-subtracted exposure was then divided by a normalized
flat-field, wavelength calibrated and fully extracted. The spectra are
flux-calibrated and telluric corrected using the IDL-based code
\verb*#xtellcor_general# described in the paper by
\citet{2003PASP..115..389V}.

The final J- and K-band spectra along with the most prominent features
are shown in Figs. \ref{fig:spectrumJ} and \ref{fig:spectrumK},
respectively.

\subsection{Equivalent Widths}

The equivalent width (EW) of six prominent features in the J
and K-band spectra were estimated using the the IDL-based function
\verb*#measure_ew# by N. Konidaris and J. Harker. Each spectrum was
normalized and the equivalent widths were calculated by computing the 
ratio of the area of a feature to a pseudocontinuum.
This pseudocontinuum is a linear interpolation of clean regions on
either side of the feature.
Uncertaities in the equivalent width were obtained using the procedure
described in \citet{1992ApJS...83..147S}.

The equivalent widths of
K  {\footnotesize I} (1.178 $\mu$m),
Fe {\footnotesize I} (1.189 $\mu$m) and
Mn {\footnotesize I} (1.290 $\mu$m) in J-band as well as 
Na {\footnotesize I} (2.206 $\mu$m \& 2.209 $\mu$m),
Ca {\footnotesize I} (2.261 $\mu$m, 2.263 $\mu$m \& 2.265 $\mu$m) and
$^1$$^2$CO(2,0) (2.293 $\mu$m) in K-band are shown in
Fig. \ref{fig:ews}. Crosses depict the stars with [M$/$H]
$<$ [M$/$H]$_{\sun}$ (GJ 544 B, GJ 611 B and GJ 166 C) and asterisks
depict the stars with [M$/$H] $>$ [M$/$H]$_{\sun}$  (GJ 324 B and NLTT
25869). A diamond represents GJ 164 AB.

\subsection{Metallicity of GJ 164}

Except for the Fe {\footnotesize I} lines that do not exhibit any
correlation with spectral type \citep{2005ApJ...623.1115C}, most
of the spectral features shown in Fig. \ref{fig:ews} are strongly
temperature-dependent \citep{1995AJ....110.2415A,
  1996MNRAS.280...77J}.
The dispersion in EW between objects with roughly the same temperature
and gravity will therefore provide insights on the metal abundances,
and the spectrum of GJ 164 shows features of strength comparable to
the metal-rich stars of our sample, especially in K-band
(cf. Fig. \ref{fig:spectrumK}).
Except for K {\footnotesize I} (1.178 $\mu$m) and Fe {\footnotesize
  I} (1.189 $\mu$m), the EW of its lines (cf. Fig. \ref{fig:ews}) make
GJ 164 more likely to be metal-rich than metal-poor.

The high value of GJ 164's $^1$$^2$CO(2,0) (2.293 $\mu$m) EW
(Fig. \ref{fig:ews}, bottom right panel) can be  attributed to its
secondary. Indeed, the data presented in Section \ref{sec:apm} shows
that the contrast ratio between GJ 164 A (M4) and GJ 164B (M8) is
close to 5:1 in K.
Since the strength of the CO bands increases with later types in the
M-Dwarf sequence, GJ 164 B (unresolved with TripleSpec) adds a
non-negligible contribution to the strength of these molecular
features.
Figure \ref{fig:mod_diff} compares the K-band spectrum of GJ 164 to
{\footnotesize PHOENIX} spectra \citep{1999ApJ...512..377H}. A
M4-Dwarf spectra alone as well as a spectrum composed of a M4
and a M8 with a 5:1 contrast ratio, all for [M$/$H]=0.0.
Although the strength of the Na {\footnotesize I} doublet and the 
Ca {\footnotesize I} triplet in the {\footnotesize
PHOENIX} spectra differ from our data, the strengths 
and shapes of the CO bands predicted by the models match the spectra
of GJ 164.
The difference between these two spectra at the
bottom of Fig. \ref{fig:mod_diff} shows that the only features
affected by the M8 companion at these wavelengths are the CO
bands.
% Similar effects can be found in the K {\footnotesize I}
%doublets in J-band, since they increase in strength from $\sim$M4 to
%$\sim$L3 \citep{2005ApJ...623.1115C}. Is this true? what it is the contrast ratio in J?, if it is the same then this is justified...

\section{MASS-LUMINOSITY RELATIONS}

The individual apparent magnitudes calculated in Section
\ref{sec:astrometry}, combined with the absolute parallax deduced from
the dynamical analysis performed in Section \ref{sec:dynaMass},
determine the following absolute H and K magnitudes:
$M_H = 7.721 \pm 0.051$ and $M_K = 7.407 \pm 0.047$ for GJ 164 A,
$M_H = 9.557 \pm 0.051$ and $M_K = 9.128 \pm 0.047$ for GJ 164 B.

\begin{figure}
\center{\resizebox*{
    \columnwidth}{!}{\includegraphics{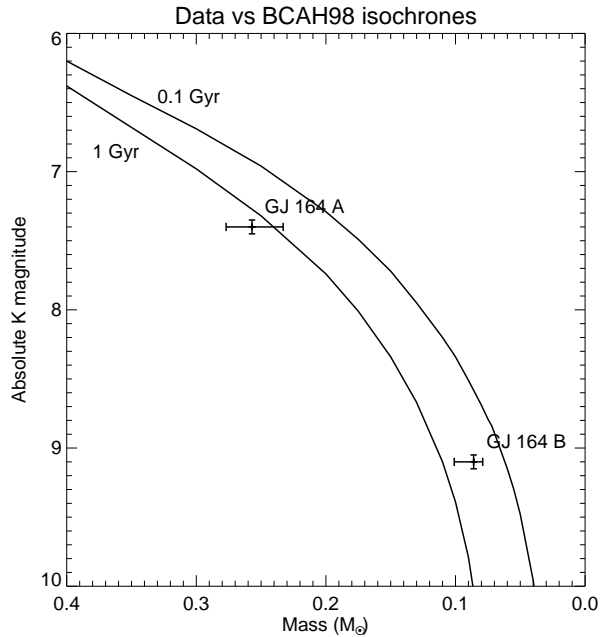}}}
\caption{
  GJ 164 A and B in a Mass-Luminosity diagram. The absolute K
  magnitudes are compared to the BCAH98 isochrones for 0.1 and 1.0
  Gyr. Asymmetric confidence intervals on the masses of both
  components are due to the uncertainty on the contrast ratio at the
  wavelength where the astrometry was taken.}
\label{fig:model}
\end{figure}

The (H-K) colors, corroborated by the infrared spectroscopy data
presented in Section \ref{sec:spectro} indicate that GJ 164 is at
least of Solar metallicity.
Figure \ref{fig:model} compares the location of GJ 164 A and B in a
Mass-luminosity diagram, relative to the low-mass Solar metallicity
models of \citet{1998A&A...337..403B} (hereafter refered to as BCAH98)
for the K-band.
For masses below 0.4 M$_{\sun}$ (but above the substellar limit),
models predict very little evolution between 0.5 and 5 Gyr.
Fig. \ref{fig:model} therefore only plots two isochrones, for 0.1 and
1 Gyr.

The aperture masking data places the primary less than 1-$\sigma$ away
from the 1 Gyr isochrone, and 3-$\sigma$ away in mass from the 0.1 Gyr
isochrone. The hypothesis of an evolved age is supported by the
kinematics of the system.
Indeed, with the proper motion (-324.5, -808.4) mas yr$^{-1}$, the
absolute parallax $\Pi = 72 \pm 1.2$ mas derived in Section
\ref{sec:dynaMass}, as well as the radial velocity $V = -29.9$ km
s$^{-1}$ measured by \citet{2004ApJ...617.1323P}, one can calculate
the Galactic space velocity $(U,V,W) = (-24,-20,-47)$ km s$^{-1}$
after correction for standard solar motion.
\footnote{The sign convention is the one of the IDL astrolib gal\_uvw
  procedure, with U positive toward the Galactic anti-center, V
  positive in the direction of Galactic rotation, and W positive
  toward the North Galactic Pole.}
When compared to the properties of the nearest young moving groups
(cf. Table 1 of \citet{2006ApJ...643.1160L}), GJ 164 exhibits a
velocity component perpendicular to the Galactic plane too large to be
associated to any of these groups.
From the kinematic properties of the M-Dwarf survey of
\citet{2002AJ....124.2721R}, it is however impossible to conclude
whether GJ 164 belongs to the thin or the thick disk.

The case of GJ 164 B is a little bit more puzzling. Indeed, as the
mass decreases, the luminosity predicted by the models becomes a very
steep function of the mass. GJ 164 B lies 1.5 $\sigma$ away from
the 1 Gyr isochrone horizontally, which is only marginally in conflict
with the models.
However, based on these dynamical mass and luminosity measurements,
the models underestimate the luminosity of a $<$0.1 M$_\sun$ star by
roughly one magnitude.
A way to accomodate for this discrepancy would be for the GJ 164 A to
be a tight binary, like for example the case of GJ 802 exposed in
\citet{2008ApJ...678..463I}. Spectroscopic observations by
\citet{2004ApJ...617.1323P} however showed no evidence that GJ 164 A
is a spectroscopic binary.

\begin{figure}
\center{\resizebox*{
    \columnwidth}{!}{\includegraphics{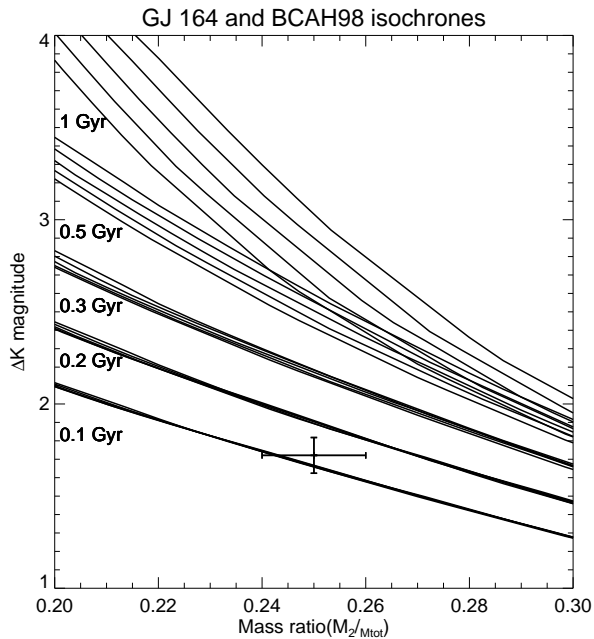}}}
\caption{
  The mass ratio and K-band contrast ratio of GJ 164 (datapoint of
  coordinates (0.25, 1.8)) are compared to BCAH98 isochrones for ages
  between 0.1 and 1 Gyr. Each age generates a family of models taking
  into account the uncertainty on the total mass of the binary.
}
\label{fig:model2}
\end{figure}

To provide a stronger case, Figure \ref{fig:model2} plots more of the
BCAH98 isochrones (0.1, 0.2, 0.3, 0.5 and 1 Gyr), this time in 
the plane mass ratio f (cf. eq. \ref{eq:astrometry}), against the
K-band magnitude difference. $\Delta$K is a direct measurement made
with the aperture masking interferometry data (cf. Table
\ref{tbl:phot}) while f is a derived product of the joined fit,
obtained in Section \ref{sec:dynaMass}. Both observables are
parallax-independent, which makes them more robust than the individual
masses and luminosities.

Although taking into account all the uncertainties of our analysis by
generating families of models within $\pm$ 1 standard deviation of the
total mass of the binary, the 1.5-$\sigma$ discrepancy between the
data and the models presented in Fig. \ref{fig:model} now turns out to
be over 5 $\sigma$ away from the 1 Gyr isochrone: the models presented
in Fig. \ref{fig:model2} clearly favor a very young age (100 Myr),
which accomodates for the apparent excessive luminosity of GJ 164 B,
but is incompatible with the kinematics of the system. Note that the
same conclusion holds when performing a similar analysis using H-band
mass-luminosity relations.

\section{CONCLUSION}

GJ 164 is an astrometric binary, whose orbit was resolved by aperture
masking interferometry observations at Palomar and Keck. The
consistency of the data is excellent, and exhibits a 1-2 mas precision
in average for each measurement, made below what is usually accepted
as the resolution limit of a telescope. This data, combined with
earlier astrometric measurements, provides the following dynamical
masses:
0.257 $\pm$ 0.020 M$_{\sun}$ for the primary and
0.086 $\pm$ 0.007 M$_{\sun}$ for the secondary.

Analysis of its infrared spectrum reveals that the M-Dwarf binary GJ
164 is at least of Solar metallicity. When attempting to use
theoretical models for Solar metallicity very low-mass stars to derive
an age from the mass and luminosity of both components, we find that
the models do not adequately fit the data, requiring very young ages
to accomodate for an overluminous secondary.
The models therefore underestimate the luminosity of very low-mass
stars that have settled on the main sequence.

The precision of the dynamical data we present here, of the order of
10 percents, however does not provide a strong constraint for the
models. 
A precise measurement of the contrast ratio in the R band, where the
astrometry data was taken would already help to better constrain the
mass ratio and ultimately improve the constraint on the individual
masses, and make our statement about the models stronger.
The very small separation (at most, 90 mas) makes this measurement
difficult, even for HST. But because it lies so close from the
substellar limit, GJ 164 B is a target that deserves most attention.
In the meantime, a Radial Velocity curve would provide an independent
measurement of the mass ratio. With an expected Radial Velocity
amplitude of 2.8 km/sec, and a period of two years, observations of GJ
164 with TripleSpec and TEDI will settle this uncertainty.

\acknowledgments
We thank the staff and telescope operators of Palomar Observatory and
Keck Observatory for their support.
B. Rojas-Ayala thanks Travis Barman for providing the {\footnotesize
  PHOENIX} models used for this work.
This work is partially funded by the National Science Foundation under
grants AST-0504874 and AST-0705085.
This publication makes use of the Simbad database, operated at CDS,
Strasbourg, France and the data products from the Two Micron All Sky
Survey, which is a joint project of the University of Massachusetts and
the Infrared Processing and Analysis Center/California Institute of
Technology, funded by the National Aeronautics and Space Administration
and the National Science Foundation.
We wish to extend special thanks to those of Hawaiian ancestry on
whose sacred mountain we are privileged to be guests. Without their
generous hospitality, the observations presented herein would not have
been possible.

\bibliography{ms}

\end{document}